\documentclass[%
 aip,
rsi,%
 amsmath,amssymb,
 reprint,%
]{revtex4-2}

\usepackage{xcolor}
\usepackage{graphicx}
\usepackage{dcolumn}
\usepackage{bm}
\usepackage{nicefrac}
\newcommand{\n}[1]{\mathbf{#1}}

\begin{document}

\preprint{AIP/123-QED}

\title[Rodr\'{i}guez et al.]{Phases and phase-transitions in quasisymmetric configuration space}

\author{E. Rodr\'{i}guez}
 \altaffiliation[Email: ]{eduardor@princeton.edu}
 \affiliation{ 
Department of Astrophysical Sciences, Princeton University, Princeton, NJ, 08543
}
\affiliation{%
Princeton Plasma Physics Laboratory, Princeton, NJ, 08540
}%

\author{W. Sengupta}
 \altaffiliation[Email: ]{ws3883@princeton.edu}
 \affiliation{ 
Department of Astrophysical Sciences, Princeton University, Princeton, NJ, 08543
}
\affiliation{%
Princeton Plasma Physics Laboratory, Princeton, NJ, 08540
}%

\author{A. Bhattacharjee}
 \altaffiliation[Email: ]{amitava@princeton.edu}
 \affiliation{ 
Department of Astrophysical Sciences, Princeton University, Princeton, NJ, 08543
}
\affiliation{%
Princeton Plasma Physics Laboratory, Princeton, NJ, 08540
}%

\date{\today}

\begin{abstract}
We explore the structure of the space of quasisymmetric configurations identifying them by their magnetic axes, described as 3D closed curves. We demonstrate that this topological perspective divides the space of all configurations into well-separated quasisymmetric phases. Each phase is characterized by the self-linking number (a topological invariant), defining different symmetry configurations (quasi-axisymmetry or quasi-helical symmetry). The phase-transition manifolds correspond to quasi-isodynamic configurations. By considering some models for closed curves (most notably torus unknots), general features associated with these phases are explored. Some general criteria are also built and leveraged to provide a simple way to describe existing quasisymmetric designs. This constitutes the first step in a program to identify quasisymmetric configurations with a reduced set of functions and parameters, to deepen understanding of configuration space, and offer an alternative approach to stellarator optimization that begins with the magnetic axis and builds outward.
\end{abstract}

\maketitle

\section{ Introduction:} \label{sec:intro}

Quasisymmetric stellarators have long been a promising magnetic confinement design concept.\cite{boozer1983,nuhren1988,tessarotto1996} Sharing the neoclassical properties of axisymmetric tokamaks\cite{boozer1983} while preserving the freedom provided by the three-dimensionality of stellarators, there have been many attempts towards designing such configurations\cite{Anderson1995,Zarnstorff2001,Najmabadi2008,Ku2010,bader2019}. However, achieving quasisymmetry (QS) throughout the volume of an equilibrium plasma was soon realized to be limited by the so-called \textit{Garren-Boozer overdetermination} conundrum\cite{garrenboozer1991b}. The conundrum emerged from the fact that when the governing equations are studied perturbatively by means of expansions in the distance from the magnetic axis, the problem becomes overdetermined beyond second order. As recently discussed, this problem results from the conflict between QS requirements and the highly restricted form of magnetohydrostatic equilibria with isotropic pressure (MHS).\cite{rodriguez2020i,rodrigGBC} \par
Even though overdetermined equations may have solutions, it was assumed, by and large, that the Garren-Boozer conundrum was symptomatic of a deeper malady---the non-existence of globally quasisymmetric MHS. This then led to a search of approximately quasisymmetric configurations through optimization in which deviations from QS were penalized.\cite{Anderson1995,Zarnstorff2001,Najmabadi2008,Ku2010,bader2019} This approach has proven to be practical in stellarator design studies but has the unfortunate appearance of being a `black-box'. Little is known (beyond experience gained by performing numerous optimization studies) of the global spatial structure of the optimization space and how different forms of cost functions shape it.\cite{rodriguez2021opt}  \par
The present paper is the first of a sequence of papers where we will attempt to shed some light on the structure of configuration space as it pertains to QS. To do so, we identify configurations approximately with sets of ordered functions and parameters, beginning with the magnetic axis and moving outward. Formally, this set corresponds to a truncated description in the so-called near-axis expansion (NAE). This simplifying approach permits an understanding of solutions, but it also offers an alternative framework in which potential designs could be sought.\par
The magnetic axis, a closed magnetic field line, and a degenerate flux surface around which nested flux surfaces accrue is an appropriate point of departure for our approach. Although one cannot attempt to understand all the relevant aspects of a configuration by just focusing on the axis, we will show that deep insight into the space of solutions can be gained by studying the axis, and in particular, some of its salient topological features. Future publications will complete the description of configurations in the framework by considering additional parameters. \par
In Section II, we introduce the basics of QS and the NAE. In Section III, the focus is placed on the magnetic axis and the interesting consequences that flow. We introduce the notion of \textit{quasisymmetric phase} and associate it to the topological self-linking number. Section IV presents three different models for magnetic axes in which the phase diagrams and associated properties are explored. Out of the analysis, several optimization-relevant observations are extracted. Section V includes some applications of the concepts in the paper, including the development of semi-quantitative criteria to assess common quasisymmetric designs. We close with some concluding remarks.

\section{Quasisymmetry and near-axis expansions} \label{sec:introQS}
Let us begin with a brief introduction to the notion of QS. Weak QS\cite{rodriguez2020} is the minimal property of the magnetic field, which grants the dynamics of charged particles an approximate conserved dynamical quantity to leading order in the gyroradius.\footnote{We simplify the picture by assuming that the electrostatic potential shares the specific QS to leading gyro-order.} This definition of QS is a single-particle concept, independent of the nature of the underlying equilibrium.\cite{tessarotto1996,burby2020,rodriguez2020} Thus, the considerations in this paper should be regarded as quite general. Succinctly, the QS condition can be written as $\nabla\psi\times\nabla B\cdot\nabla(\mathbf{B}\cdot\nabla B)=0$, where $2\pi\psi$ is the toroidal flux, $\mathbf{B}$ is the magnetic field and $B=|\mathbf{B}|$. A QS configuration possesses nested flux surfaces\cite{burby2019,rodriguez2020,rodriguez2021isl}, specified by a well-defined $\psi$. \par
The formal definition of QS in its triple vector form above, although succinct, is not the most convenient for constructing solutions. It is more helpful to rewrite the conditions in an explicit coordinate-dependant form. This can be generally achieved by using a special set of straight field line coordinates: generalized Boozer coordinates (GBC)\cite{rodrigGBC,rodriguez2020i}. These coordinates are an extension of Boozer coordinates, which coincide with them only when the field satisfies $\mathbf{j}\cdot\nabla\psi=0$, where $\mathbf{j}$ is the current density defined by $\mathbf{j}=\nabla\times\mathbf{B}$. Using GBC, it can be shown that the QS condition reduces to the requirement that the magnetic field magnitude $B=B(\psi,\chi=\theta-\tilde{\alpha}\phi)$, where $\{\psi,\theta,\phi\}$ are the coordinates and $\tilde{\alpha}$ is an integer. \par
To complete the description of the quasisymmetric field (including the specification of GBCs) we must satisfy the following two so-called magnetic equations\cite{rodriguez2020i,rodrigGBC,garrenboozer1991a} (not to be confused with magnetic differential equations\cite{newcomb1959}),
\begin{align}
    (B_\alpha(\psi)-&\Bar{\iota}B_\theta)\frac{\partial\n{x}}{\partial\psi}\times\frac{\partial\n{x}}{\partial\chi}+B_\theta\frac{\partial\n{x}}{\partial\phi}\times\frac{\partial\n{x}}{\partial\psi}+B_\psi\frac{\partial\n{x}}{\partial\chi}\times\frac{\partial\n{x}}{\partial\phi}=\nonumber\\
    &=\frac{\partial\n{x}}{\partial\phi}+\Bar{\iota}(\psi)\frac{\partial\n{x}}{\partial\chi}, \label{eq:co(ntra)variant}
\end{align}
\begin{equation}
    \frac{B_\alpha(\psi)^2}{B(\psi,\chi)^2}=\left|\frac{\partial\n{x}}{\partial \phi}+\Bar{\iota}\frac{\partial\n{x}}{\partial \chi}\right|^2. \label{eqn:Jgen}
\end{equation} 
In the equations above, GBCs have been employed explicitly as independent variables, most notably in the partial derivatives. Doing so has introduced the position vector $\mathbf{x}$ explicitly into the equations. To proceed further, we need to define $\mathbf{x}$. It is convenient to do so with respect to the magnetic axis. The magnetic axis, described by $\mathbf{r}_0$, is a closed 3D curve around which the toroidal nested flux surfaces are constructed. The position vector can then be written,\cite{garrenboozer1991a,landreman2018a,rodriguez2020i}
\begin{equation}
    \mathbf{x}=\mathbf{r}_0+X(\psi,\theta,\phi)\hat{\mathbf{\kappa}}+Y(\psi,\theta,\phi)\hat{\mathbf{\tau}}+Z(\psi,\theta,\phi)\hat{\mathbf{b}}. \label{eqn:xEq}
\end{equation}
The unit vectors represent the Frenet-Serret basis associated with the axis. In this form, the position vector can describe points in space parametrized by GBCs. \par Equations~(\ref{eq:co(ntra)variant})-(\ref{eqn:xEq}) constitute a highly coupled set of nonlinear partial differential equations (PDEs). Given their complexity, a preferred theoretical approach \cite{Mercier1964,Solovev1970} is to consider this problem perturbatively using NAE. This approach is concerned with the solution around the axis, which makes the definition of the displacement $\mathbf{x}$ with respect to $\mathbf{r}_0$ convenient. Tackling the construction of solutions from the inside out is opposite to the common approach of finding equilibrium magnetic field solutions, whereby one specifies the boundary and solves equations \textit{inwards}, with the axis emerging as part of the solution. This salient difference between the old approach and the new, which elevates the role of the magnetic axis, will enable us to develop a simpler and ordered understanding of configurations and ways to optimize them. \par
In our perturbative expansions of the equations, the expansion parameter is taken to be a quasi-radial coordinate constructed from the toroidal flux $\epsilon=\sqrt{\psi}$ (where the normalisation constant is taken to be unity for simplicity). One can then express functions as Taylor (in $\epsilon$)-Fourier (in $\theta$) series,\cite{garrenboozer1991a,landreman2019,rodriguez2020i}
\begin{equation}
    f=\sum_{n=0}^\infty\epsilon^n{\sum_{m=0|1}^{n}}\left[f_{nm}^c(\phi)\cos m\chi+f_{nm}^s(\phi)\sin m\chi\right].
\end{equation} 
The PDEs then reduce to an ordered set of ordinary differential and algebraic equations. We do not repeat the resulting equations here, which have been discussed elsewhere\cite{rodriguez2021weak}.  \par
To set up the resulting problem governes by those equations consistently, one needs to provide as inputs certain parameters and functions at every order in the expansion.\cite{rodriguez2021weak,rodriguez2020i,landreman2019} This essentially identifies solutions/configurations (at least approximately) with ordered sets of inputs (see Table~\ref{tab:QSconfigCharParam}). A truncated set then serves as a \textit{model} with a particular geometric meaning for many configurations that share some common properties and allows us to explore the structure of configuration space systematically.\par  
In the present paper, we identify configurations exclusively by the shape of the magnetic axis, which, as shown in Table~\ref{tab:QSconfigCharParam} constitutes the $0^\mathrm{th}$ order description. A more complete description requires additional choices of parameters and will be presented in future publications. For instance, if one focuses on vacuum QS configurations that are stellarator symmetric, then the scalar parameters $\eta$ and $B_{22}^C$ need to be carefully chosen to build a second-order description, which includes several essential features of relevant configurations.
\begin{table}[]
    \centering 
    \begin{tabular}{|c|c|}
        \hline
    Order & Params. \\ \hline\hline
    $0$ & $B_0$, axis ($\kappa,~\tau,~l)$ \\
    $1$ & $B_{\theta 20}$, $\sigma(0)$, $\eta$ \\
    $2$ & $B_{22}^C$, $B_{22}^S$, $B_{\alpha1}$ \\\hline
    \end{tabular}
    \caption{\textbf{Quasisymmetric configuration characterising parameters.} The table gathers the free parameters (and functions) defining the leading order form of quasisymmetric configurations. In this paper we focus on the 0th order axis shape and leave a discussion of the remainder to a future publication.}
    \label{tab:QSconfigCharParam}
\end{table}

\section{Zeroth order: magnetic axis} \label{sec:0th}
\par

\subsection{Describing the magnetic axis} \label{sec:axisDef}
The magnetic axis is a three-dimensional closed curve \textit{central} to the stellarator. Up to an Euclidean isometry (i.e., a rotation or translation), such curves are uniquely identified by their curvature $\kappa$ and torsion $\tau$ (the fundamental theorem \cite{eisenhart1909treatise} of curves for non-vanishing $\kappa$ ). These functions $\kappa$ and $\tau$ are also the forms in which the axis is represented in the near-axis formalism discussed in Sec.~\ref{sec:introQS}. Thus, it seems natural to attempt to understand the space of magnetic axes (and thus of quasisymmetric configurations) by studying the different forms that the curvature and the torsion might take. However, the choice of curvature and torsion is highly nontrivial, as a general choice will result in a curve that does not close on itself. (No simple analytical form of curvature and torsion that produce closed curves \cite{fenchel1951differential} is known to exist.) \par
To guarantee that the 3D curve is closed, it is convenient to describe the position of its loci directly. Requiring the curve to close then becomes tractable: closure of the axis is equivalent to periodicity in the parametrization of the curve. However, the price to pay for tractability is that we lose direct control on the functions $\kappa$ and $\tau$. \par
To fix ideas, we define the axes explicitly as the loci $\mathbf{r}_0=R(\phi)\hat{R}+Z(\phi)\hat{Z}$, where $\hat{R}$ and $\hat{Z}$ are the cylindrical unit vectors and
\begin{align}
    R=1+\sum_{n=1}^N\left[R_n^C\cos(nN_\mathrm{nfp}\phi)+R_n^S\sin(nN_\mathrm{nfp}\phi)\right], \label{eqn:axisDefR}\\
    Z=\sum_{n=0}^N\left[Z_n^C\cos(nN_\mathrm{nfp}\phi)+Z_n^S\sin(nN_\mathrm{nfp}\phi)\right]. \label{eqn:axisDefZ}
 \end{align}
Here $N_\mathrm{nfp}\in\mathbb{N}$ represents the number of field periods (or equivalently the toroidal $N$-fold symmetry) of the axis, and $\phi$ is the cylindrical angle. We normalize all lengths to the major radius. Parametrizing closed curves in this form is common practice (see, e.g., the equilibrium solver VMEC\cite{hirshman1983}). However, we note that this parameterization already excludes a whole family of curves such as curves that turn on themselves, i.e., curves with lengths that are not a monotonic function of $\phi$. Equations~(\ref{eqn:axisDefR})-(\ref{eqn:axisDefZ}) do however include self-intersecting curves (which are a set of measure zero).  \par
For simplicity, let us further restrict ourselves to the set of stellarator-symmetric configurations, i.e., those for which the system is symmetric under the discrete mapping $\phi\rightarrow-\phi$ and $Z\rightarrow-Z$. This parity-like symmetry is commonly adopted in the design of stellarators mainly due to the resulting reduction of the parameter space and simplification of geometry (including coil sets that inherit the symmetry).\cite{dewar1998} In the parametrization provided, stellarator symmetry is imposed straightforwardly taking $R_n^S=0$ and $Z_n^C=0$. Our axes are then represented by sets of harmonics $\{R_n, Z_n\}$.
In Sections~\ref{sec:undersConfSpace} and \ref{sec:qualAssessAxis}, we discuss the role and meaning of these harmonics using some representative families.

\subsection{Quasisymmetric phases} \label{sec:QSphases}
Two essential features of quasisymmetric configurations can be deduced entirely from those of the magnetic axis. These features hold for any weakly quasisymmetric configuration and are thus quite general. First, the curvature of the magnetic axis must not vanish anywhere. Breaking this requirement would imply that, to preserve QS, flux surfaces around the axis would become infinitely elongated ribbons.\cite{garrenboozer1991b,landreman2018a,rodriguez2020i} This naturally separates the space of configurations into those that have non-vanishing curvature on the magnetic axis and curves with \textit{inflection points}\cite{aicardi2000} (where $\kappa=0$). This requirement on the regularity of the curves guarantees the existence of the Frenet-Serret frame. \par
In this frame, there exists a normal vector, $\hat{\kappa}$, defined everywhere along the axis. As noted in [\onlinecite{landreman2018a}], the number of times that the normal vector encircles the axis directly corresponds to the symmetry of the configuration. This is often called the helicity of QS and is defined by the integer $\tilde{\alpha}$ introduced in Section \ref{sec:introQS}. This integer distinguishes quasi-axisymmetric (QA) from quasi-helically symmetric (QH) configurations. While QA is defined by $\tilde{\alpha}=0$ (toroidal symmetry), closest analogue to axisymmetry, QH is defined by $\tilde{\alpha}\neq0$. (From the near-axis perspective, quasi-poloidal symmetry, in which the direction of symmetry is purely poloidal, is prohibited.\cite{plunk2019}) \par
We present a brief explanation underlying this geometric interpretation of the symmetry. From the NAE, it follows that to leading order $X_{11}^C=\eta/\kappa>0$ (see Eq.~(14) in [\onlinecite{rodriguez2020i}], assuming $\eta>0$). Now, at some constant $\phi$ cross-section, consider the point where the magnetic field magnitude is maximum to first-order in the expansion. This corresponds to $\chi=0$, and thus $\mathbf{x}-\mathbf{r}_0=\epsilon\eta/\kappa\hat{\mathbf{\kappa}}+\epsilon Y_1\hat{\mathbf{\tau}}$. Given $X_1>0$, this position vector never deviates by more than $\pi/2$ away from $\hat{\mathbf{\kappa}}$. Thus, if the normal vector encircles the magnetic axis $\tilde{\alpha}$ times, then so must the contour corresponding to the maximum of $|\mathbf{B}|$. It follows that the rotation of the normal and the QS helicity are identical.\footnote{These considerations are as general as weak QS is. However, the argument can be made simpler in the case of equilibria with $\mathbf{j}\times\mathbf{B}=\nabla p$. In that case, the gradient of $B$ on the axis is along the normal $\nabla B=\kappa B \hat{\kappa}$. Hence, the motion of the normal with that of contours of $B$ follows again.} \par
It is thus clear that the shape of the axis completely determines the helicity of the symmetry.\cite{landreman2018a,rodriguez2020ii} Each regular closed curve can then be assigned a form of QS. To do so, it is necessary to count the number of turns of the normal. In practice, a possibility is to evaluate the area swept by the normalised curvature vector in the plane perpendicular to the axis as one completes a toroidal turn. Defining the normalised curvature vector $\hat{\kappa}=(\kappa_R^0,\kappa_\phi^0,\kappa_Z^0)$ in the cylindrical basis, we obtain
\begin{equation}
    \tilde{\alpha}=\frac{1}{2\pi}\int_0^{2\pi}\left[\kappa_Z^0\frac{\mathrm{d}\kappa_R^0}{\mathrm{d}\phi}-\kappa_R^0\frac{\mathrm{d}\kappa_Z^0}{\mathrm{d}\phi}\right]\mathrm{d}\phi. \label{eqn:helCompQS}
\end{equation}
Here the sign convention has been taken to be consistent with the definition of $\chi=\theta-\tilde{\alpha}\phi$. Under this convention, if we move in the $+\phi$ direction of the integral, the curvature vector would rotate anti-clockwise and have a positive sign. \par
\begin{figure}
    \centering
    \includegraphics[width=0.45\textwidth]{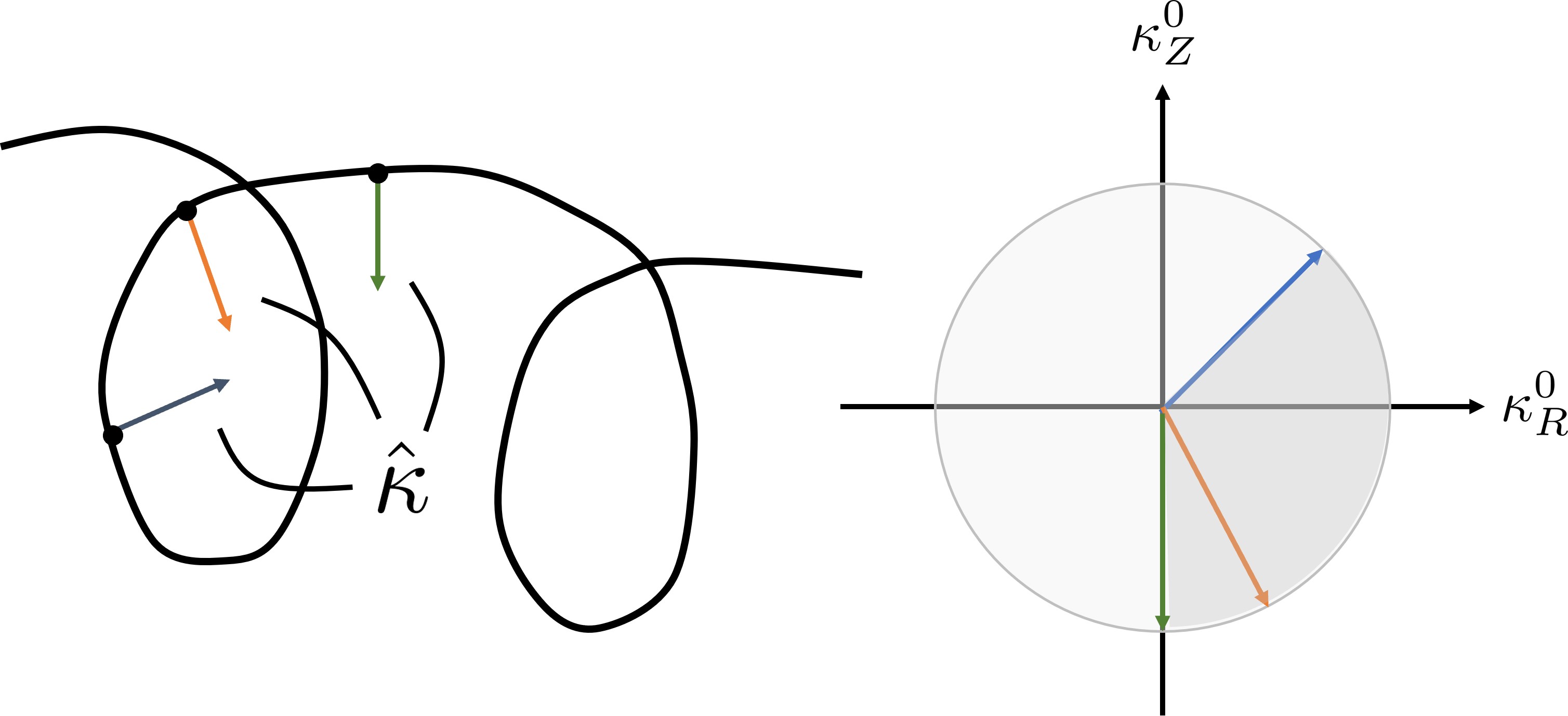}
    \caption{\textbf{Schematic of the calculation of helicity $\tilde{\alpha}$}. Diagram illustrating the calculation of $\tilde{\alpha}$ and its geometric meaning. The left shows the curvature vector at three points along the axis, and the right shows the swept area.}
    \label{fig:alphConstructInt}
\end{figure}
The expression given by Eq.~(\ref{eqn:helCompQS}) is exact. However, it leaves opaque how the QS-helicity manifests itself in configuration space. Although we are certain that each regular curve is associated with a value of $\tilde{\alpha}$, the expression does not elucidate the structure of the space. Are configurations sharing $\tilde{\alpha}$ naturally grouped or scattered in the space?  \par
To make progress, we connect the helicity of QS to the topological notion of \textit{self-linking number} ($SL$).\cite{fuller1999,oberti2016,moffatt1992} The $SL$ is an extension of the common concept of linking number between curves. It is defined as the linking number between the axis and the curve generated by displacing the axis along the direction normal to the axis. This is precisely the same concept as $\tilde{\alpha}$. The power of this association is that $SL$ is a topological invariant under regular isotopies, i.e., continuous deformations of the closed curve that do not pass through intermediate curves with vanishing curvature. This means that the space of all closed curves is separated into regions identified by an integer value of $SL$. Each of the pockets shares the same symmetry, and to change that symmetry, one must go across a manifold made up of curves possessing inflection points. \par
Such a transition manifold corresponds to the non-quasisymmetric configurations mentioned at the beginning of this subsection. Although these axes do not serve as quasisymmetric configurations, they can be identified with a different class of optimized stellarators, namely \textit{quasi-isodynamic} configurations.\cite{plunk2019} Such configurations are approximately \textit{omnigeneous} (i.e., on average prevent particles from collisionlessly drifting radially), with closed $|\mathbf{B}|$ contours in the poloidal direction. Their on-axis magnetic field is not constant, unlike in the case of QS. The turning points of $B_0$ match the points at which the curvature of the axis vanishes. The number of vanishing curvature points along the axis then determines the number of trapping wells on the axis. Given that this class of optimized stellarators is identified with the transition manifolds, it might appear at first glance that the space of quasi-isodynamic configurations is much reduced. However, this is not so: the additional freedom on $B_0(\phi)$ and other parameters make the space of \textit{quasi-isodynamic} configurations a large one.
\par
The analysis based on $SL$ is powerful because it enabled us to dissect the whole space of configurations (identified by their axes) into well-separated regions. To the class of configurations that each region represents, we shall refer to as a \textit{quasisymmetric phase}, characterized by a particular value of the self-linking number. To change the symmetry of the configuration by continuous deformations, one necessarily needs to cross the \textit{phase-transition} manifold, where no QS solutions exist. This reinforces the view that quasisymmetric phases are distinct solution classes that generally possess quite distinct properties. It also sheds some light on efforts that navigate the space of solutions searching for QS. If restricted to the space of quasisymmetric configurations, the symmetry of the starting point will restrict the solutions found to that particular phase. We note that traditional optimization is seldom performed in that space. However, once an optimizer manages to reduce the measures of QS to a low enough value, it is effectively confined to one of these phases. For instance, starting from a tokamak or a rotating ellipse with a circular axis will likely bias optimization towards the QA phase. If a QH phase is found, it is likely biased towards the transition with QA. There is, in that sense, a memory of the starting point in optimization. 

\section{Understanding configuration space} \label{sec:undersConfSpace}
In the previous Section, we have introduced the crucial, topological concept of \textit{quasisymmetric phase} and the realization that the space of configurations is divided into well-distinguished regions corresponding to different forms of QS. This concept is helpful, but still somewhat abstract (it is hard to identify it directly to a closed curve as described by Eqs.~(\ref{eqn:axisDefR})-(\ref{eqn:axisDefZ})). It would also be illuminating to understand the interaction between the form of the QS phase and other properties of these configurations. In what follows, we consider some representative models. Understanding these models will help elucidate qualitative trends that can be applied to more general situations.
\par
\subsection{Symmetric torus unknot} 
Let us start with the simplest of cases, one that is characterized by a single harmonic. We write
\begin{subequations}
\begin{gather}
    R=1+a\cos N\phi, \\
    Z=a\sin N\phi.
\end{gather}
\end{subequations}
The family of curves spanned by $\{a, N\}$ is commonly known as \textit{torus unknots}\cite{oberti2016} and can be pictured as closed helices with a circular cross-section. The properties of this family of closed curves have been thoroughly studied in other contexts.\cite{fuller1999,oberti2016,aicardi2000}. \par
\begin{figure}
    \centering\hspace*{-0.4cm}
    \includegraphics[width=0.5\textwidth]{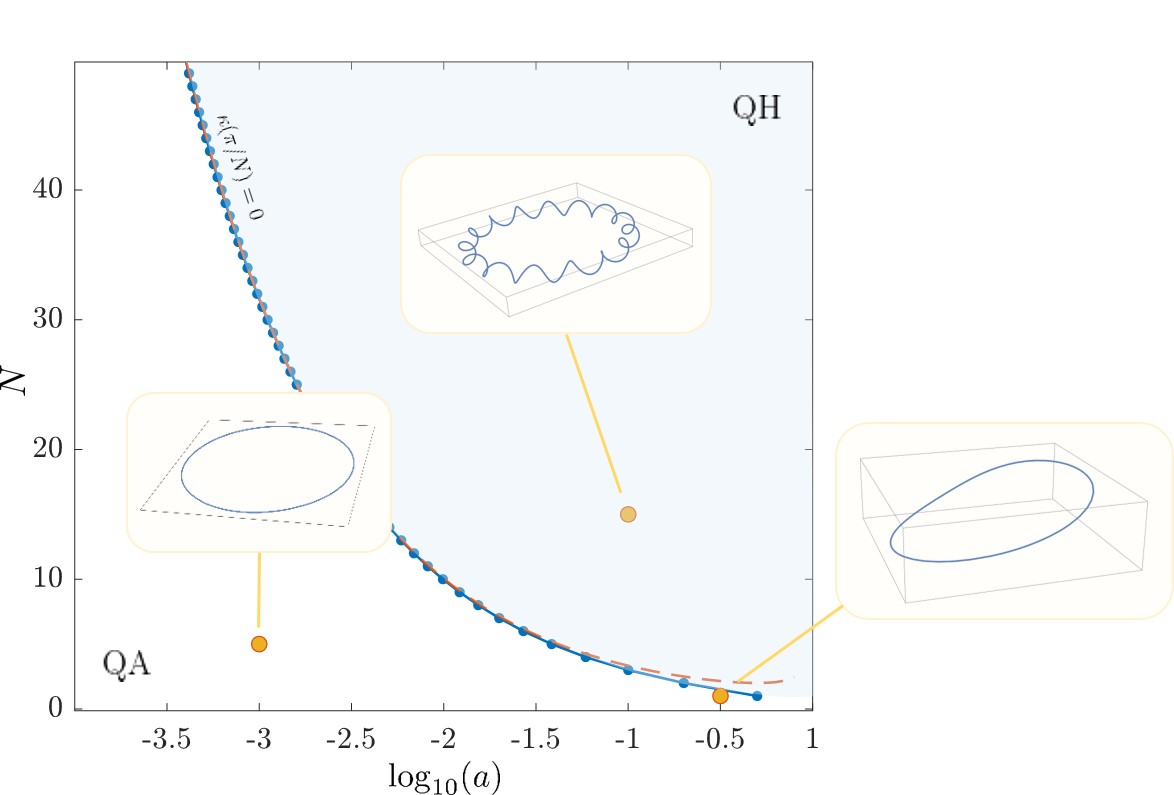}
    \caption{\textbf{Space of symmetric torus unknots.} The space of symmetric torus knots spanned by $\log_{10}(a)$ and $N$ is presented. The two QS phases are indicated (QA and QH, shaded blue) separated by the curve at $a=a_\mathrm{crit}$. Three explicit illustrating examples of curves are given. The broken orange curve represents the $N\sim 1/\sqrt{a(1-a)}$ estimate.}
    \label{fig:QAQHphaseTorusKnot}
\end{figure}

We now explore the structure of the configuration space spanned by these two parameters. The space is divided into two phases (see Figure \ref{fig:QAQHphaseTorusKnot}), separated by a single phase transition boundary\cite{fuller1999,oberti2016}. For $a<1$ and given $N$, the boundary corresponds to a critical $a=a_\mathrm{crit}$, choice that leads to the curvature vanishing at $\phi=\pi/N$. The analytic expression for the phase-transition is
\begin{equation}
    a_\mathrm{crit}=\frac{1}{1+N^2}. \label{eqn:aCritTorKnot}
\end{equation}
Defining the measure $s=a(1+N^2)$, a value $s>1$ corresponds to QH with $\tilde{\alpha}=-N$, while $s<1$ leads to a QA. One may show this either numerically (applying Eq.~(\ref{eqn:helCompQS})) or analytically (see [\onlinecite{oberti2016}]). The measure $s$ provides a relevant scale for the shaping of the axis. \par
The division of the space into these two phases depending on the value of $a$ has a clear geometric interpretation. Let us think of the magnetic axis as a combination of a large circle (of unit radius) and a helix (of radius $a$ and pitch $2\pi/N$). The total curvature (and thus the direction of the normal to the axis) can be regarded as a result of the competition between these two contributions. When the pitch of the helix is small, the curvature of the unit radius circle dominates, and the normal vector does not rotate (QA phase). As the helix coils more (because the number of turns increases with $N$ or $a$ increases), the curvature vector will tend to rotate with the helix. In the latter case, we recover the QH phase. This approximate geometric picture can be used to estimate the phase transition as the point where the curvature of the helix equals that of the circle. This gives $N\sim 1/\sqrt{a(1-a)}$, which agrees in the limit of large $N$ (or small $a$) with the exact form of $a_\mathrm{crit}$. \par
The simple torus unknot model makes explicit the well-known fact that QHs require more significant shaping than QAs do, not just in the form of the excursion of the axis $a$, but also in the number of field periods $N$ (see the measure $s$). As a matter of fact, more field periods are necessary for the axis excursion of the QH phase to be moderate compared to the QA phase. For example, for $a$ below 10\% of the major radius, one is forced to consider $N\geq4$, a typical value in the design of QH configurations. The observation that a larger number of field periods is natural to the QH phase implies that $\phi$ derivatives of various quantities are also naturally enhanced by an extra factor of $N$. This is especially important for higher orders in the near-axis construction, as they depend on derivatives of lower-order quantities. Qualitatively, in an expansion dominated by these derivatives, the number of field periods can be thought of as effectively reducing the expansion parameter $\epsilon$, $\epsilon_\mathrm{eff}\sim N\epsilon$.
\par
Let us now return to the phase transition. Recall that the boundary is made up of axes incompatible with QS (although they may be identified with \textit{quasi-isodynamic} solutions)\cite{plunk2019}. Configurations at the transition formally possess surfaces that look like infinitely elongated ribbons. However, this physically unrealizable elongation does not only apply to the transition itself but also affects the region around it. This can be easily seen from the approximate expression for elongation ($\mathcal{E}$, the ratio of major and minor axes of the elliptical surfaces), obtained by near-axis expansion\cite{rodriguez2020ii,landreman2018a,garrenboozer1991b}, $\mathcal{E}\sim Y_{11}^S/X_{11}^C\sim2\sqrt{B_0}(\kappa/\eta)^2$. A vanishingly small curvature leads to strong shaping of the surfaces. Note that the first-order parameter $\eta$ clearly and directly affects the shaping of the surfaces of the configuration, and thus a full description of the shaping is not within the scope of this paper. However, regardless of the choice of $\eta$, some shaping features, such as the large elongation near the transition, are unavoidable.  \par
In point of fact, strongly $\phi$-dependent curvature configurations will always exhibit nontrivial surface shaping. As presented in Appendix A, the variation is strongest close to the transition. In the QH phase, the $\phi$-variation is minimum away from the transition at approximately $a\sim1/N$ (see Appendix A for more details). Minimizing the variation in $\phi$ also appears to be appropriate for reducing large derivatives and minimizing the size of higher-order near-axis contributions.   \par
Upon first glance, the discussion above suggests that the region close to the transition should be avoided when choosing QS designs. In the QA phase, then, one could think that making the shaping as small as possible would be preferred in designs. However, this extremum corresponds to the axisymmetric limit, and doing so implies losing much of the appeal of the stellarator. To prevent this, we shall bring the on-axis rotational transform to the picture. Although calculations of the on-axis rotational transform involve features of the shaping of the surrounding surfaces once again, there is an essential geometric contribution from the shape of the axis. Such contributions (which have been related to the Berry phase (or Hannay's angle)\cite{bhattacharjee1992}) can be seen in Mercier's expression for the rotational transform (see [\onlinecite[Eq.~(44)]{Helander2014}]). The integrated torsion and $SL$ are two important contributors, and these can be combined into a property of the axis: the \textit{writhe}. This is a property of every closed curve that can be interpreted as an `intrinsic' rotational transform, as discussed in [\onlinecite[Eq.~(62)]{pfefferle2018}]. In a way, the writhe provides a measure of the potential contribution of the axis to $\Bar{\iota}$. As shown in Appendix B (where further details and analysis of the writhe may be found), the writhe increases with $a$. Consequently, to grant a significant geometric contribution to $\Bar{\iota}_0$, the configurations of interest in the QA phase should not lie too far from the phase transition (and, in fact, not have too large $N$). In the QH phase, the opposite is true: further away from the transition, the contribution is largest (effectively bounded by $N$).\par
In the QH phase, though, the further away from the transition one looks, the larger the excursions of the axis are. Excessive coiling of the axis is detrimental when it limits the effective aspect ratio of the resulting configuration. This occurs if two non-sequential segments of the axis lie very close to one another so that the surfaces around each segment would touch each other. The placement of coils to generate these fields would also become challenging. We show a quantitative account of this \textit{self-distance} in Appendix C, to exclude in practical terms the region in the QH phase roughly beyond $a\gtrsim1/N$. \par
Thus, in the QA phase, we have learned that an optimal region for the design of configurations lies close to the transition, but not too much due to a competition between the geometric contribution to the rotational transform and the excessive shaping of surfaces. On the other hand, in the QH phase, the optimal region lies somewhere between the transition and $a\sim1/N$, where the shaping of both surfaces and the axis is appropriate.
\par

\subsection{Elliptic torus unknot}
In this and the following subsections, we explore the generalization of the phase diagrams discussed above by looking at extended families of closed curves. This will enable us to understand how the different harmonic contributions to Eqs.~(\ref{eqn:axisDefR})-(\ref{eqn:axisDefZ}) compete and shape the configuration space. \par
Let us first extend the family of symmetric torus unknots by extending the circular cross-section to an elliptic one. The three-parameter family for the axis then reads,
\begin{subequations}
\begin{gather}
    R=1+a\cos N\phi, \\
    Z=b\sin N\phi.
\end{gather}
\end{subequations}
Of course, $a=b$ corresponds to the case of symmetric torus unknots that we have considered already. To understand how the structure of the configuration space is modified for $b\neq a$, we once again find the loci of configurations with vanishing curvature. Not unsurprisingly, a QA-QH phase transition occurs at $a_\mathrm{crit}=1/(1+N^2)$ (independent of $b$). This manifold corresponds to vanishing curvature at $\phi=\pi/N$, leaving the transition boundary between QA and QH intact. \par
Beyond this transition, as one changes $b$, curves do change shape, and one might expect additional phase changes. Take, for instance, the limit of $b\rightarrow0$ in the QH-phase. In the limit, the curve becomes planar, and, therefore, cannot have a rotating axis normal. It can be shown that such curves possess inflection points and thus constitute a phase transition. In this case, a transition between QH phases with opposite helicities (corresponding to a reversal of the handiness of the helical pitch). In the QA phase, there is no transition. \par
\begin{figure*}
    \centering
    \includegraphics[width=\textwidth]{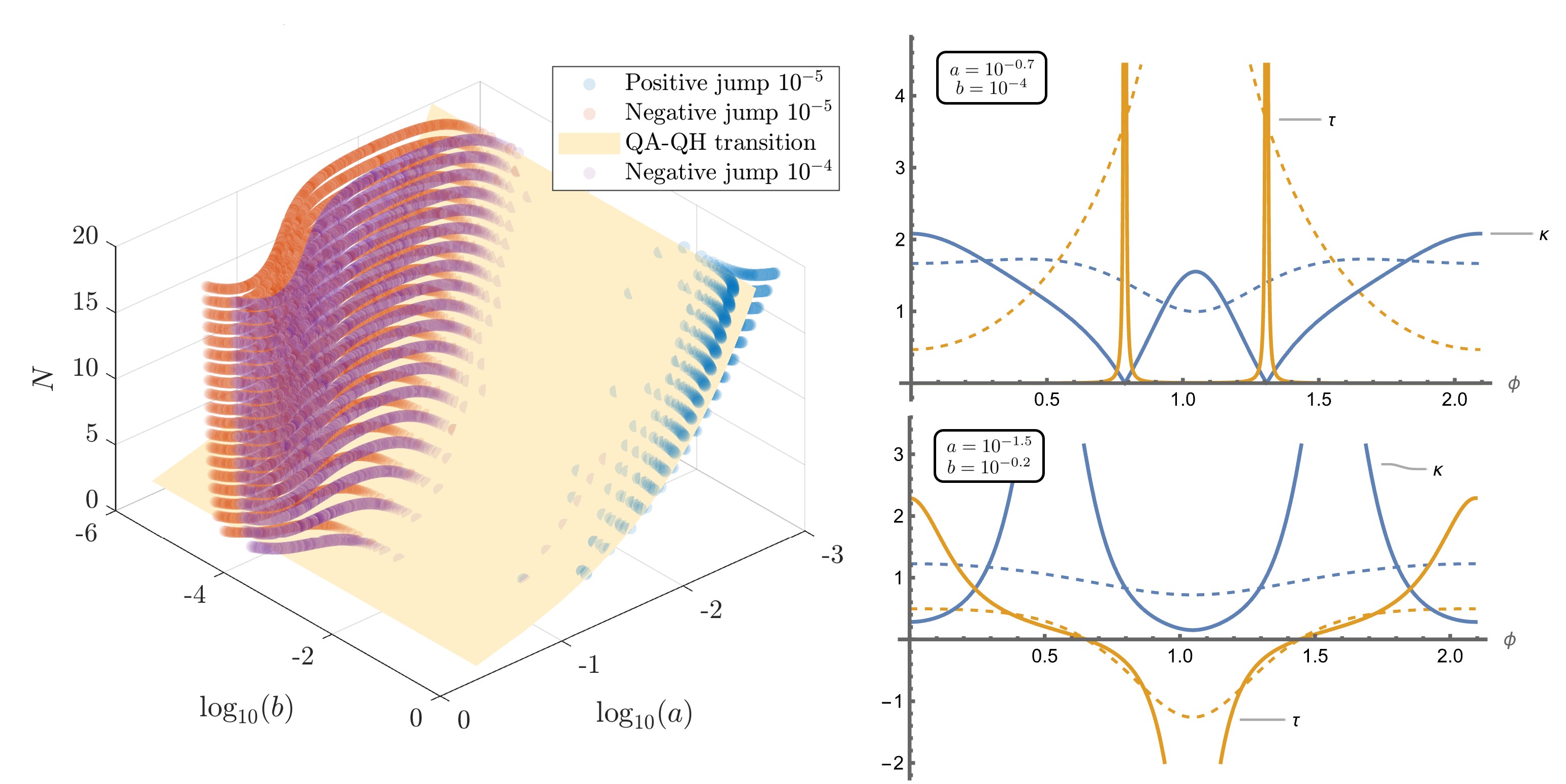}
    \caption{\textbf{QS phase space for elliptic torus unknots.} The space of elliptic torus knots is presented as a function of $\{a,b,N\}$, with the QA-QH transition surface. We also plot as scatter points configurations for which the minimum curvature corresponds to $10^{-4}$ and $10^{-5}$. The plots in the right correspond to the curvature and torsion in the interval $(0,2\pi/3)$ for a QH (top) at $b$ close to 0 ($\log_{10}(a)=-0.7$, $\log_{10}(b)=-4$, $N=3$), and QA (bottom) for large $b$ ($\log_{10}(a)=-1.5$, $\log_{10}(b)=-0.2$, $N=3$). The broken lines represent the curvature and torsion for $a=b$.}
    \label{fig:torusKnotEllip}
\end{figure*}

Away from the transition, in the QH phase, there is no other in $b$. This is so even though the geometry of the QH configurations is significantly affected by $b$. In fact, the curvature becomes increasingly close to $0$ at some points along the axis as the $b=0$ transition is approached (linearly with $b$), as seen in Fig.~\ref{fig:torusKnotEllip}. As noted previously, a small curvature leads to an unrealistic ribbon-like shaping of surfaces. Thus, the low curvature resulting from the decrease in $b$ suggests that, effectively, the space of QHs is reduced to $b$ values not too dissimilar from $a$. A quantitative form for this variation is not easily obtained. However, the scaling of the minimum curvature with $b$ for small $b$ can be used as a measure. \par
The effects of $b$ on the QA side are not as dramatic. Lower $b$ values leave the curvature largely unaffected, and one may show that for $a(1+N^2)\ll 1$ and large $N$, to leading order the curvature $\kappa\approx1+a(2+N^2)\cos N\phi$ for $b=0,~a$ (although $\tau$ does significantly change). For large $b$ values, for which the axis excursions become large, significant curvature variations occur, but one does not encounter additional transitions. As an example of this behavior, consider the curvature at $\phi=\pi/N$. In the QA side, assuming $b\sim1$, $\kappa(\pi/N)\sim[1-a(1+N^2)]/(1+N^2)$. Note that the effects are most noticeable near the transition boundary (as the numerator is small). \par
We show the structure of the 3D phase space in Figure \ref{fig:torusKnotEllip}. We should note that when including additional harmonics to the description of the axis (which comes now), we can expect some of this invariance with $b$ to prevail. However, the significant changes in the geometric contribution of the harmonics could also affect some of the phase structures when these arise from a competition between multiple harmonic contributions. We shall, however, focus on the case where $R_n\sim Z_n$ for the following sections.

\subsection{Anharmonic torus unknots} The last model we will explore considers the interaction of two different harmonics describing the axis. The family of curves is now written as,
\begin{subequations}
\begin{gather}
    R=1+a\cos N\phi+b\cos kN\phi, \\
    Z = a\sin N\phi +b\sin kN\phi.
\end{gather}
\end{subequations}
Here $k$ is an integer. How is the QS-phase space modified due to this additional competing harmonic? In the limit of a single dominant harmonic (i.e., a symmetric torus unknot), we already know that the QA-QH transition boundary must occur at $a\sim1/(1+N^2)$ (or $b\sim1/(1+k^2N^2)$ if $a=0$). Thus, we expect the space to be partitioned into at least three distinct phases: a QA, a QH with helicity $N$, and a QH with helicity $kN$. We now show that the different harmonics can also interact to result in additional phases. \par
Consider the domain of $a$ and $b$ to span both positive and negative values (i.e., to span $[-1,1]$). We do this because the relative sign between the harmonics could be important. One might guess that the contributions to helicity may add or subtract depending upon the sign combinations. Of course, not all sign combinations are independent, as a rotation by $\pi/N$ leads to a map of the form
\begin{align*}
    R\rightarrow 1-a\cos N\phi+(-1)^kb\cos Nk\phi, \\
    Z\rightarrow -a\sin N\phi+(-1)^k\sin Nk\phi.
\end{align*}
This means that for an even $k$, the configuration space must be symmetric about $a=0$, as $a\rightarrow-a$ must leave the space unchanged. The symmetry is different for odd $k$, as in that case, the symmetry is through the origin, and the space is invariant under the operations $a\rightarrow-a$ and $b\rightarrow-b$. \par
Finding the transition manifolds in this model to construct the phase portrait is much more complicated than before. In fact, only some of those transitions can be simply described analytically. We shall resort to numerical evaluation for the others. The main reason is that inflection points $\kappa=0$ appear at movable $\phi$ values due to various geometric effects competing. Let us then first focus on the transitions that can be found exactly. \par
Looking for zeroes of curvature at $\phi=0$ and $\pi/N$, the transition manifolds are 
\begin{subequations}
\begin{gather}
    1+a(1+N^2)+b(1+k^2N^2)=0, \\
    1-a(1+N^2)+(-1)^kb(1+k^2N^2)=0.
\end{gather} \label{eqn:anharQAQHsurf}
\end{subequations}
These define two lines in $\{a,b\}$ configuration space. They are the extension of the symmetric torus unknot transition (and include these). Most notably, the product $s_a=a(1+N^2)$ is once again the relevant measure that shapes the space of QS phases. These transitions are remarkable in that they are \textit{exact} and \textit{universal}, as they can be straightforwardly generalized for any number of harmonics and regardless of $Z_n$ choice (see Appendix \ref{sec:invarPhaseTrans}). 
\par
\begin{figure*}
    \centering
    \includegraphics[width=\textwidth]{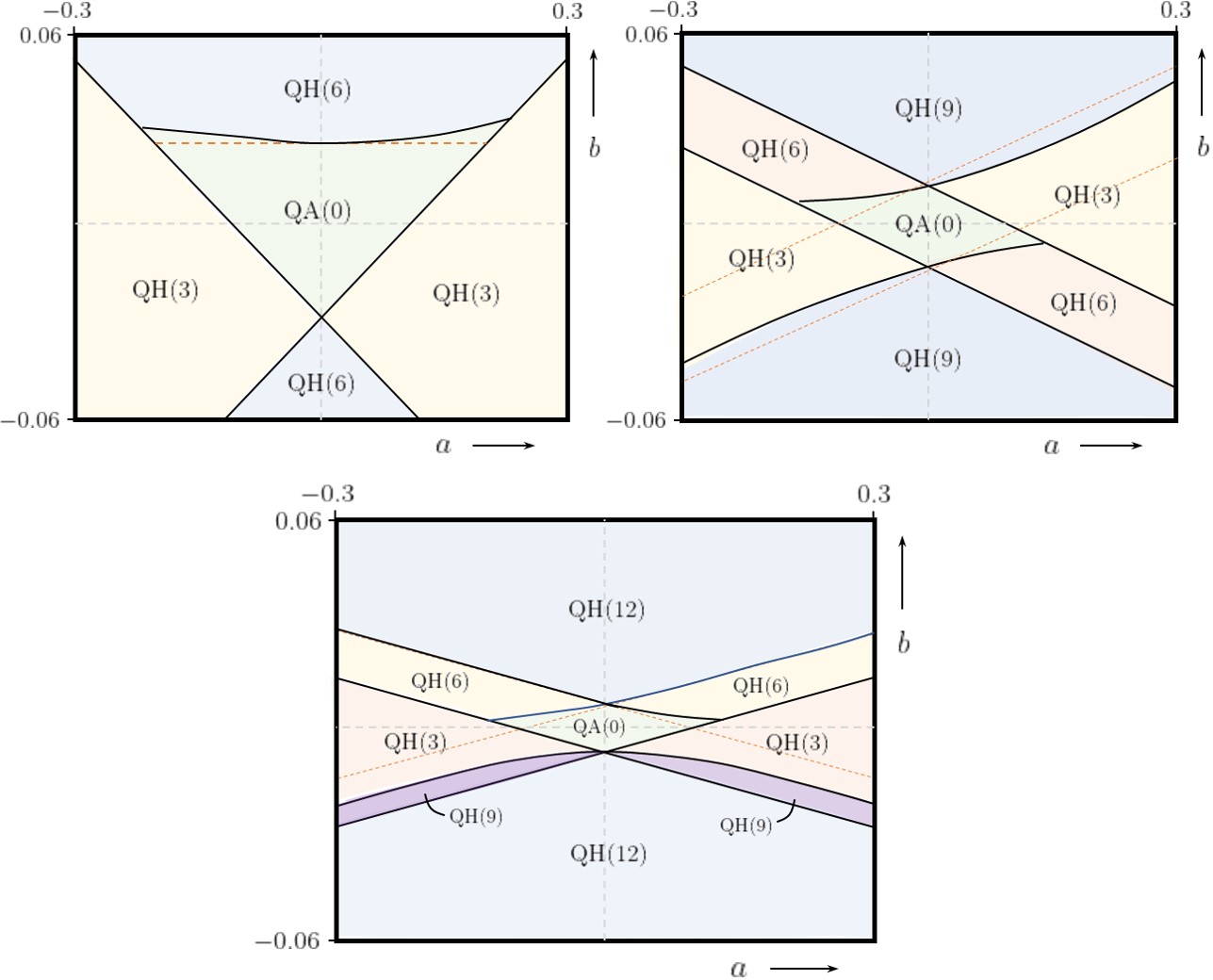}
    \caption{\textbf{QS phase space for anharmonic torus knots for $k=2,~3,~4$.} The space of anharmonic torus knots is presented as a function of $\{a,b\}$, for $N=3$, for a number of harmonic values $k=2,~3,~4$ (clockwise). The different phases are found numerically, with some of the estimated boundaries presented as broken orange lines. Each phase is labeled according to its helicity. The diagrams are representative of all number of field periods $N$ if one represents the axes in $s$ rather than absolute values.}
    \label{fig:torusKnotAnharks}
\end{figure*}

The phase space contains additional transitions beyond these universal transitions and the symmetric torus unknot points. The complexity of the structure is illustrated in Figure \ref{fig:torusKnotAnharks} for $k=2,~3,~4$ and $N=3$. There we can identify the transitions of Eq.~(\ref{eqn:anharQAQHsurf}), but also transitions related to the complex interaction between the geometric contributions of the different harmonic. This complexity makes it not only harder to find closed-form descriptions for these transitions but also makes them more susceptible to change. We noted the universality of the transitions in Eq.~(\ref{eqn:anharQAQHsurf}) with the number of field periods (when using the measure $s_a$) or the relative magnitude of $R_n$ and $Z_n$, but we expect these other transitions to change with them. We shall not explore these further here.   \par
Let us analyse the $k=2$ case which will turn out to be particularly relevant. The configuration space is separated into five distinct regions (corresponding to QS phases with three helicity labels). A central region of low $b$ and $a$ corresponds to QA, and opposite QH regions correspond to $N$ and $2N$. In this case, we may separate the phases according to the following criterion. A QA state corresponds to the region for which $s_a$ and $s_b=b(1+4N^2)$ roughly satisfy $s_b\leq1$ and $s_a-s_b-1<0<s_a+s_b+1$. A QH state is realized with the helicity corresponding to the $b-$harmonic for $s_b>1$ and $s_a-s_b-1<0<s_a+s_b+1$, or $s_a-s_b-1>0>s_a+s_b+1$. The helicity-$N$ QH phase corresponds to the remaining configuration space, that is, $(s_a-s_b-1)(s_a+s_b+1)>0$.
\par
More generally, for a given $k$, we expect to find $k+1$ distinct QS-helicity phases. One may find $\tilde{\alpha}=0,N,\dots,kN$, with two disjoint $kN$ regions in the irreducible symmetry cell for even $k$. This means that the number of interfaces must also grow. Given the symmetry requirements in the configuration space, we expect $k+1$ surfaces in the irreducible sector of configuration space. Of these surfaces, we know with precision two, given by (\ref{eqn:anharQAQHsurf}), and the transition points along the $b=0$ and $a=0$ lines. \par
\par
We draw attention to the broken orange lines in Figure \ref{fig:torusKnotAnharks}, which include some estimates for the transitions. In particular, in addition to the expressions for (\ref{eqn:anharQAQHsurf}), we have considered the other two sign combinations. For $k\geq3$, these are not too bad as guides for the configuration phase. (Here, we have not derived a general expression and cannot expect our analysis to work in all scenarios.)
\par
The variety of transitions also leads to a wider variety of quasi-isodynamic configurations with different number of trapping magnetic field wells. The number of inflection points along the magnetic axis can be related to the difference in the helicities of the bounding quasisymmetric phases.\cite{aicardi2000} This means that (assuming we have simple inflection points, that is, curvature vanishes only locally) at a given transition surface, the curve will possess several vanishing curvature points equal or greater to the change in $\tilde{\alpha}$ across the transition. In fact, one may check this on the phase space presented in Fig.~\ref{fig:torusKnotAnharks}, where the lower bound is satisfied.   \par
This intricate structure exhibits part of the complexity achieved when multiple harmonics coexist. It also makes it harder to predict \textit{a priori} what a given axis would be without computing its properties explicitly. Although this complexity is patent, it also seems that the phases $0,~N$, and $kN$ are the prominent ones. As one moves between one and the other, one traverses several phase transitions between $N$ and $kN$, where the harmonics interact to generate intermediate helicities. Generally, it seems that these phases are restricted to relatively narrow regions in configuration space. This tends to make these configurations strongly shaped because the curvature on the axis might be close to zero at several points. Such configurations should be avoided. \par

\section{Practical application and qualitative assessment} \label{sec:qualAssessAxis}
Exploring the models considered above has given us some quantitative and qualitative perspectives on the importance and implication of the shape of magnetic axes. Can we use them to assess qualitatively the case of a general axis described by (\ref{eqn:axisDefR}) and (\ref{eqn:axisDefZ})? An axis described by a large set of coefficients $\{R_1,R_2,\dots,Z_1\dots\}$ will produce a wide variety of phases and geometric properties \textit{ala} Fig.~\ref{fig:torusKnotAnharks}. In practice, we would want to have a way to judge which harmonics are relevant and what are reasonable values for them. Of course, the answer will depend on the properties of the configuration we are interested in. This is clear within the NAE approach, as it is not the same to inquire about the kind of phase the axis represents (a 0th order question) or the potential QS quality (a second-order question). Higher orders tend to amplify the contribution of higher harmonics due to the presence of recurrent derivatives so that even the lowest coefficients can become relevant. \par
With this in mind, let us start by analyzing the contribution to the QS phase. The primary tool to gauge the relevance of harmonics is by comparing the magnitude of the product $s_n=R_n(1+n^2)$ to unity and each other. We have taken only the radial harmonics in constructing this metric, assuming that $Z_n\approx R_n$. This is, in practice, a reasonable simplifying choice (backed by the behavior of the elliptic torus unknot, invariance of many features of configuration space, and existing designs). However, it must be borne in mind that significant disparity in $R_n$ and $Z_n$ can lead to important changes in the phase diagrams. Configurations violating this assumption can be studied as a different class of designs, and we do not consider them further here.  \par
If none of the $\{R_n\}$ meets the criterion $s_n>1$, then the axis will likely be QA. Of course, if all measures are $s_n\ll1$, the phase reduces to axisymmetry. To avoid this, finding configurations with at least one $s_n$ close to unity is of interest. We can test this criterion through a practical example. Take the magnetic axis from the QA design in [\onlinecite{landreman2021}] (the so-called precise QA) obtained directly from the VMEC files, and write the leading three harmonics: $R_{\{2,4,6\}}=\{0.184,~0.0217,~0.00260\}$ and $Z_{\{2,4,6\}}=\{0.158,~0.0206,~0.00256\}$. First, we note that, as we hypothesized, $R_n\sim Z_n$. The QS measure $s_n=\{0.92,0.37,0.10\}$, which clearly agrees with a QA with a close-to-unity harmonic. \par
If at least one $s_n$ is greater than unity, we expect to be a QH. When a single harmonic dominates, the symmetric torus unknot is expected to be a good model, and we expect to find $s_n$ somewhere between unity and $n$. In practice, though, one finds multiple harmonics to be relevant, which leads to a growth in complexity (see Figure \ref{fig:torusKnotAnharks}). Although the comparison of $s_n$ with unity is still a good first test, a more detailed comparison will be necessary using the analytic conditions (\ref{eqn:anharQAQHsurf}) or their generalisation (\ref{eqn:genAnharTrans}). Generally, we shall not need to deal with all the complex, intermediate phases, as strong shaping tends to make them impractical. Using the precise QH in [\onlinecite{landreman2021}] as a testbed, $R_{\{4,8,12,16\}}=\{0.200,0.0314,0.00467,0.000559\}$ and $Z_{\{4,8,12,16\}}=\{0.179,0.0286,0.00431,0.000528\}$. The measure $s$ is $\{3.4,2.0,0.7,0.1\}$. We expect to find a QH configuration in this case, given that at least two $s_n>1$. To study this particular combination, we need to evaluate the conditions $f_-=1+a(1+N^2)+b(1+k^2N^2)$ and $f_+=1-a(1+N^2)+b(1+k^2N^2)$, which in this case $f_-=6.4>0$ and $f_+=-0.4<0$. This takes us into the QH($N$) phase. So we have a QH configuration dominated by the $N$ harmonic, but with a significant influence from a second one. The configuration lies close to the transition. \par
\begin{figure}
    \centering
    \includegraphics[width=0.475\textwidth]{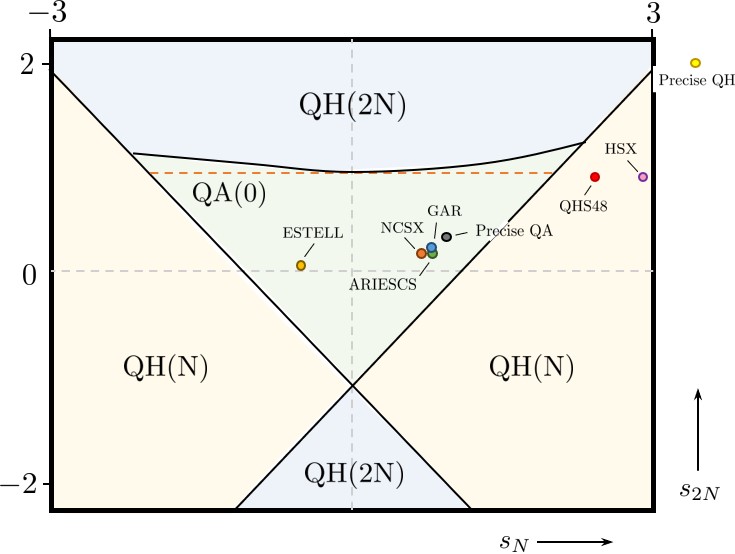}
    \caption{\textbf{Existing QS designs in the normalised $s_n$ space.} The figure shows existing QS designs as they lie in the phase diagram in which the axes represent $s_N$ and $s_{2N}$: GAR\cite{Garabedian2008,Garabedian2009}, HSX\cite{Anderson1995}, NCSX\cite{Zarnstorff2001}, ESTELL\cite{Drevlak2013}, QHS48\cite{Ku2010}, ARIESCS\cite{Najmabadi2008}, Precise QA and Precise QH\cite{landreman2021}. The phase transitions taken as reference correspond to $N=3$, $k=2$ and $R_n=Z_n$ (see Fig.~\ref{fig:torusKnotAnharks}).  }
    \label{fig:designsPhase}
\end{figure}

The construction of the dimensionless $s_n$ is a powerful step, as the behavior of phase transitions is dominated by it. To illustrate this further, we can collect existing QS designs and represent them in $s_n$ space. This gives us a unifying simple analytic principle to organize highly disparate QS configurations (including those with a different number of field periods). To represent them together, we consider the two dominant harmonics of the magnetic axes for these designs (noting that $Z_n\sim R_n$ is a good assumption), construct $s_n$, and represent them together in a 2D configuration space. We do so in Figure \ref{fig:designsPhase}, where the phase diagram for $n=3$, $k=2$ and $R_n=Z_n$ is shown as a reference. A two-harmonic space has been chosen as this is appropriate for the majority of the designs (see Table ~\ref{tab:designs}), which have small third harmonic contributions. Some (e.g., precise QH and QHS48) do show significantly higher shaping, and to better contextualize these, one would have to study a 3D equivalent to Fig.~\ref{fig:designsPhase} which includes their third harmonic explicitly, which we shall not do here. The number of relevant harmonic contributions to the phase can be judged by $s_n$ and used to group QS stellarators into different classes (denote them by a roman numeral indicating the number of harmonics, see Table \ref{tab:designs}). \par
From this representation, several observations can be made. First, QA designs seem all to be clustered in a small region in the QA phase (note that the diagram is symmetric about $s_N=0$). The QH designs, in turn, seem to be more dispersed in space, though relatively close to the phase transition of QH(N) in the $s_{2N}>0$ part of the diagram. The distribution of these designs in this space is interesting but hard to understand or predict. A more complete analysis, including some of the higher-order parameters and additional shaping, is needed for this, and we shall do that in future publications. However, Fig.~\ref{fig:designsPhase} does suggest that new unexplored potential designs might exist, as there are large areas without designs that remain to be explored.
\begin{table*}[t]
\begin{tabular}{c||cccccccc|}
    & GAR & HSX & NCSX & ESTELL & QHS48 & ARIESCS & Precise QA & Precise QH \\ \hline
    $N$ & 2 & 4 & 3 & 2 & 4 & 3 & 2 & 4 \\\hline
    $s_N$ & 0.77 & 2.9 & 0.74 &	-0.54 &	2.4 & 0.83 & 0.92 & 3.4 \\
    $s_{2N}$ & 0.29 & 0.97 & 0.19 & 0.11 & 0.97 & 0.20 & 0.37 & 2.0 \\
    $s_{3N}$ & 0.075 & 0.024 & 0.018 & -0.018 & 0.22 & 0.027 & 0.096 & 0.68 \\\hline
    Class & II & II & II & II & III & II & II & III \\
    \hline
\end{tabular}
\caption{\textbf{Dominant harmonic content of magnetic axes of designs in Fig.~\ref{fig:designsPhase}}. Table including the number of field periods, the three dominant harmonic description of the magnetic axes and class of the QS designs in Figure \ref{fig:designsPhase}. The third harmonic should serve as a measure of how well the phase diagram in Figure \ref{fig:designsPhase} captures the designs. Both QHS48 and precise QH could be considered to have some significant $3N$ shaping.}
\label{tab:designs}
\end{table*}
\par
In Fig.~\ref{fig:designsPhase} we considered the two dominant harmonic contributions. This dimensional reduction is appropriate for most of the practical designs in question as to their third harmonic contributions are generally small (see Tab.~\ref{tab:designs}, showing that most belong to Class II). A natural question arises, then: do we lose anything relevant in the description of the configuration by dropping `small' harmonics $s_n\ll1$? The brief answer is yes, and this is simply a result of the QS phase not being a complete account of the problem. The precise form of the geometry is also important, especially to determine properties of the configurations, such as the quality of QS. These properties correspond to higher orders in the NAE, meaning that higher derivatives of axis quantities are naturally involved. This amplifies the contribution of the higher harmonics. With this in mind, we could, at order $k$, propose the tentative measure $s_n^k=R_n(1+n^2)n^k$ to assess the relative importance of the harmonics that make up the axis. This can make harmonics which are judged unimportant for phase considerations, become relevant. Thus, each point in Fig.~\ref{fig:designsPhase} can be thought to represent a group of configurations with additional harmonics (small in the QS-phase sense). This added flexibility can be leveraged to find designs with additional higher order properties within a given stellarator class. This flexibility comes at a loss of simplicity, but will prove important in future publications. \par
The construction of $s_n^k$ is only an informed guess beyond the 0th order and thus must be taken as such. However, it can provide a valuable tool to understand the bounds and relevance of various harmonics. For example, for the precise QA, one needs to keep at least two harmonics for the truncated rotational transform in the model to be within 10\% of its actual value (corresponding to keeping a term with $s_4^1/s_2^1\sim0.8$). In the precise QH, the third harmonic becomes relevant as $s_{12}^1$ becomes more than half of the dominant contribution. Taking the designs in Fig.~\ref{fig:designsPhase} as a reference, it appears that construction through second-order to capture the relevant features will need to include roughly four to five harmonics. Of course, truncating this still retains crucial insight, as we have seen here and will be explored in the future.

\section{Conclusion}
In this paper, we have explored the space of quasisymmetric configurations by identifying them with their magnetic axes. Doing so corresponds to describing configurations by a model emerging from the leading order of a near-axis expansion. This is only a first step in an attempt to understand the entire QS configuration space, for which other parameters in the expansion beyond the shape of the axis are required. An analysis of these and their implications and potential for alternative optimization space will be presented in future publications. \par
Analyzing axis shapes, we have separated the space of all quasisymmetric configurations into well-defined \textit{quasisymmetric phases}. Each of these phases is characterized by a topological invariant, the \textit{self-linking number} ($SL$), corresponding to the helicity of the symmetry. One may not smoothly deform the axes and change the configuration phase without crossing a phase transition at which the configuration cannot be quasisymmetric. Such transition manifolds correspond to quasi-isodynamic axes. \par
By exploring three models revolving around the torus unknot, we show characteristic features of these phases and the structure of phase diagrams. We construct a criterion that estimates the importance of the harmonics describing the magnetic axis in this phase diagrams, the scalar $s_n=R_n(1+n^2)$. Larger values lead to more significant contributions and relative comparison to unity, roughly the QA or QH nature of the configuration. Important trends are investigated associated with these phases. This provides an approximate way to classify and discuss existing QS designs, which we present in their simplest forms. \par
Higher harmonics become more relevant when properties of the configuration are considered that pertain to higher-order NAE description, e.g., the quality of QS. An analysis of such properties will thus generally require a larger space of configurations to be considered. The importance of higher harmonics for some order $k$ in the expansion can be estimated by a modified $s_n^k$ criterion, which includes an additional $k$-th power of $n$ factor. \par

\par
\hfill

\section*{Acknowledgements}
The authors gratefully acknowledge discussion with Matt Landreman and Rogerio Jorge. This research is primarily supported by a grant from the Simons Foundation/SFARI (560651, AB) and DoE Contract No DE-AC02-09CH11466. ER was also supported by the Charlotte Elizabeth Procter Fellowship at Princeton University. 

\section*{Declaration of interests}
The authors report no conflict of interest.

\section*{Data availability}
The data supporting this study's findings are available from the corresponding author upon reasonable request.

\appendix

\section {Curvature of symmetric torus unknot}
\begin{figure*}
    \centering
    \includegraphics[width=\textwidth]{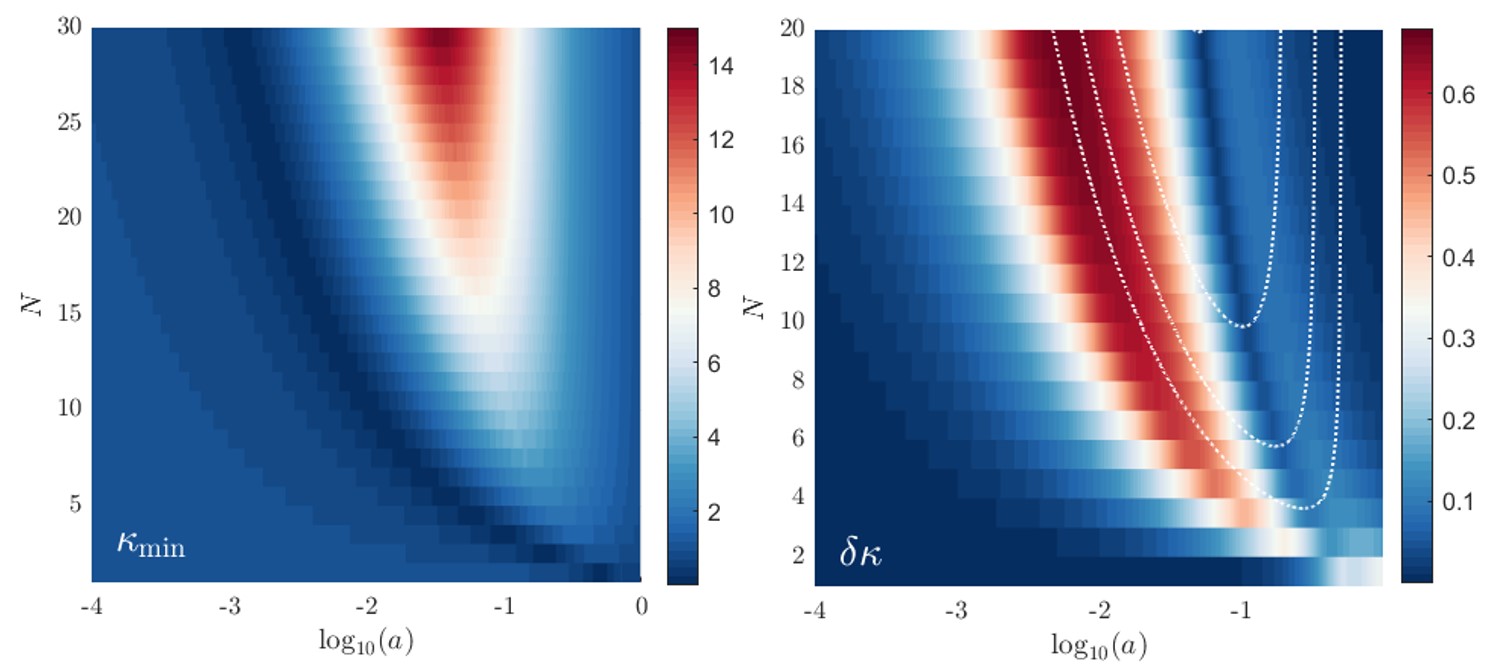}
    \caption{\textbf{Space of symmetric torus unknots for $\kappa_\mathrm{min}$ and $\delta\kappa$.} The space of symmetric torus unknots spanned by $\log_{10}(a)$ and $N$ is presented, and the respective minimum curvature value (left) and curvature variation along the axis (right) are shown. The white contour on the right represent contours of $\kappa_\mathrm{min}$ for reference.}
    \label{fig:torusKnotCurvature}
\end{figure*}

In the main text, we discussed some general aspects associated with the curvature as a function of parameter space. We explore these in more detail here. To do so, we start with Figure \ref{fig:torusKnotCurvature}, where the value of $\delta\kappa=\mathrm{std}[\kappa(\phi)]$ and $\kappa_\mathrm{min}$ are plotted in $\{a,N\}$ space. \par
We begin by considering several limits. In the limit of $a\rightarrow 0$, one recovers axisymmetric configurations. Therefore, the curvature must correspond to that of a circle with a unit radius. In particular, this implies $\kappa_\mathrm{min}\rightarrow 1$ and $\delta\kappa\rightarrow0$. The QA phase inherits this behavior, generally showing slow variations and a curvature close to unity. One may easily estimate how the variation $\delta\kappa$ grows with $a$ and $N$ in the QA phase away from the axisymmetric case. The key is to rescue the geometric picture in which the axis is composed of a unit radius circle and a helix. For a given $N$, in the limit of small $a$, one may show that the curvature of the helix $\kappa_H\sim aN^2$. Thus, we expect the variation in $\delta\kappa$ to scale like this as the helical curvature modulates $\kappa=1$ associated with the major radius. \par
In the opposite limit, that is, for a sizeable helical excursion $a\rightarrow1$ (and large $N$), we expect the curvature to be dominated by the helical curvature instead. In fact, $\kappa_\mathrm{min}\sim 1/a$ in this limit of the QH phase, as the axis becomes a collection of concatenated circles of radius $a$. As a result, $\delta\kappa\rightarrow0$ in that limit. The behavior between these two simple limits is, however, not monotonic. The QH phase shows a local minimum for the variation of the curvature and a maximum for $\kappa_\mathrm{min}$ somewhere between the transition and the limit above. This more complex structure is clear in Figure \ref{fig:torusKnotCurvature}. \par
The local minimum (valley) for $\delta\kappa$ corresponds to the maximum value of $\kappa_\mathrm{min}$. This can be understood as follows. If the minimum of the curvature is pushed up while its maximum value is bounded, then the curvature becomes more tightly packed, reducing its variation in $\phi$ to a minimum. However, this does not tell us anything about where one finds the minimum $\delta\kappa$. To do so, it is convenient to introduce the notions of \textit{normal} ($\kappa_n$) and \textit{geodesic} ($\kappa_g$) curvature.\cite{fuller1999} For our symmetric torus unknots, one may think of the geodesic curvature as the curvature of the axis \textit{on} the surface of the torus on which it lives, while $\kappa_n$ corresponds to the curvature perpendicular to it. For large $N$ and $a$, $\kappa_g^2\ll\kappa_n^2$, as the axis is mainly coiled normal to the torus. Assuming $\kappa_n$ to dominate in the majority of the QH phase, we obtain 
\begin{equation}
    \kappa\approx\kappa_n=\frac{(1+a\cos N\phi)\cos N\phi+a N^2}{(1+a\cos N\phi)^2+N^2a^2}.
\end{equation}
Then, one may find that the minimum of the expression in $\phi$ has its largest value at $a=1/N$, for which $\kappa_n^\mathrm{extr}\sim N/2$. This is in excellent agreement with the valley in Figure \ref{fig:torusKnotCurvature} except at low $N$ (where the assumption of dominant $\kappa_n$ fails). \par
The variation in curvature is maximal in the QH phase close to the transition. This also explains why the QA phase has increased variability as the transition is approached. To avoid significant variations, one would generally like to stay away from the transition. On the QH side, $a\sim1/N$ seems a convenient choice. We expect the strong shaping near transitions to remain valid for more general axis models.

\section{Writhe and `geometric' rotational transform}
The shape of the axis plays a vital role in determining the value of the on-axis rotational transform. From Mercier's form for $\iota_0$ [\onlinecite[Eq.~(44)]{Helander2014}], we know that properties of the axis such as torsion or $SL$ contribute to the transform. However, the relation is more complex and goes beyond the axis, requiring next-order NAE considerations\cite{rodriguez2020i,landreman2018a,garrenboozer1991b}. \par
\begin{figure*}
    \centering
    \includegraphics[width=\textwidth]{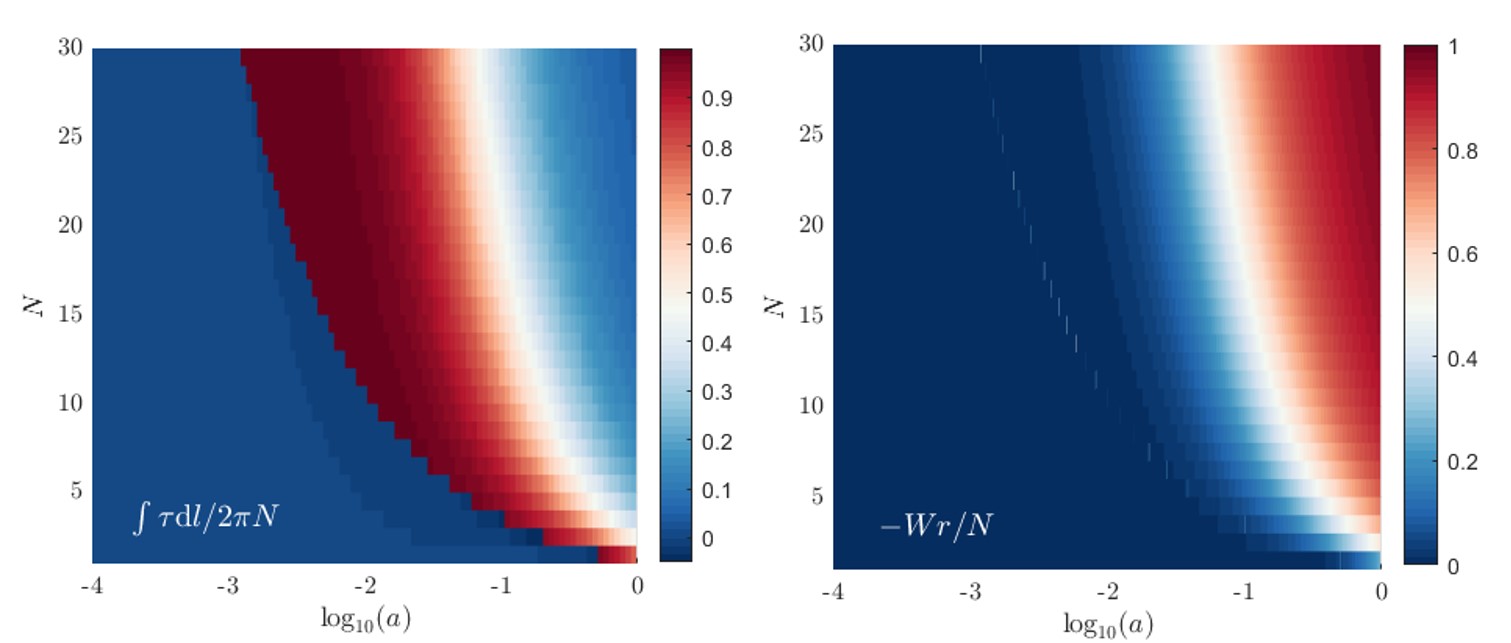}
    \caption{\textbf{Space of symmetric torus unknots for integrated torsion and writhe.} The space of symmetric torus knots spanned by $\log_{10}(a)$ and $N$ is presented, and the corresponding integrated torsion and writhe shown as a colormap. The quantities are normalised to the number of field periods. The QA phase has a negative integrated torsion. }
    \label{fig:torusKnotTorsion}
\end{figure*}

As mentioned above, both the integrated torsion and $SL$ play an important role. Both of these contributions may be agglomerated into a closed curve measure known as the \textit{writhe} (see Figure \ref{fig:torusKnotTorsion}).
The writhe may be defined through the C\"{a}lug\"{a}reanu–White–Fuller’s theorem \cite{moffatt1992,oberti2016} as 
\begin{equation}
    Wr = SL+\mathcal{T}, \label{eqn:CWFth}
\end{equation}
where $\mathcal{T}$ is the total integrated torsion, and $Wr$ is the so-called writhe. The writhe constitutes a measure of the coil and kinkiness of the curve, and as considered in [\onlinecite[Eq.~(62)]{pfefferle2018}], it represents naturally the geometric contribution to the transform on axis. Thus the additional physical meaning of the writhe. \par
The writhe undergoes no discrete jump as one crosses the phase transition, and therefore we expect the contribution to the rotational transform to vary smoothly across the transition. The same is not true of the integrated torsion, as can be seen in Fig.~\ref{fig:torusKnotTorsion}. From Eq.~(\ref{eqn:CWFth}), the change in $\mathcal{T}$ must then be the same as in $SL$ (up to a sign). \par
The trends of the writhe can provide us with critical insight into the behavior of the rotational transform in each of the QS phases. Figure \ref{fig:torusKnotTorsion} shows that the writhe increases globally with both $N$ and $a$, seemingly monotonically. We should look at these carefully, however. Let us consider looking at a fixed $N$. The increase and decrease in integrated torsion in the QA and QH phases with larger $a$ respectively explain the $a$ dependence. Torsion increases away from the axisymmetric configurations in the QA phase and dwindles away from the transition in the QH phase, where a discrete jump leads to its largest value. As a result, the rotational transform of QAs will be the largest closest to the transition, while QHs (driven by $SL$, which roughly sets the maximum attainable transform) tend to have a larger transform away from it. \par
Globally, we also observe that the writhe grows with $N$. However, we must be careful in interpreting this observation. Perhaps seemingly contrary to this trend (which is true), the value of the writhe nearby the QS transition decreases like $\sim1/N$ with $N$ after peaking for $N=2$. This can be seen from the leading-order expressions,
\begin{equation}
\mathcal{T}=\begin{cases}
        N, & a>a_\mathrm{crit} \\
        0, & a<a_\mathrm{crit},
    \end{cases}    
\end{equation}
with $O(1/N)$ corrections. Thus, this suggests that larger field periods do not enhance the transform on axis for the QA configurations. Contrary to this behavior close to the transition, deep in the QH phase, towards $a\rightarrow1$, the rotational transform will grow linearly with $N$. In fact, in the limit $a\rightarrow1$, $\mathcal{T}\sim1$ and $Wr\sim -N+1/N$. This shows the difference in the nature of the two different phases. For QH devices, rotational transform is naturally larger than that of QAs, driven by the larger helicity of the symmetry. Thus, one generally needs to be less concerned about making the value of $\iota_0$ non-vanishing when choosing a particular realization of the QH phase. When the number of field periods is low ($N\sim1,~2$), these trends, obtained in the limit of large $N$, change. Then the precise value of $\iota_0$ constitutes truly a higher-order consideration, and our exploration here is of limited value.

\section {Self-distance of the axis}
As pointed out in the main text, it might seem \textit {a priori} that there is no drawback to increasing $a$ towards $a=1$ in the QH phase. The rotational transform contribution increases and the $\phi$ dependence does not dramatically increase. However, a significant limitation arises as the axis becomes more strongly shaped. \par
\begin{figure*}
    \centering
    \includegraphics[width=\textwidth]{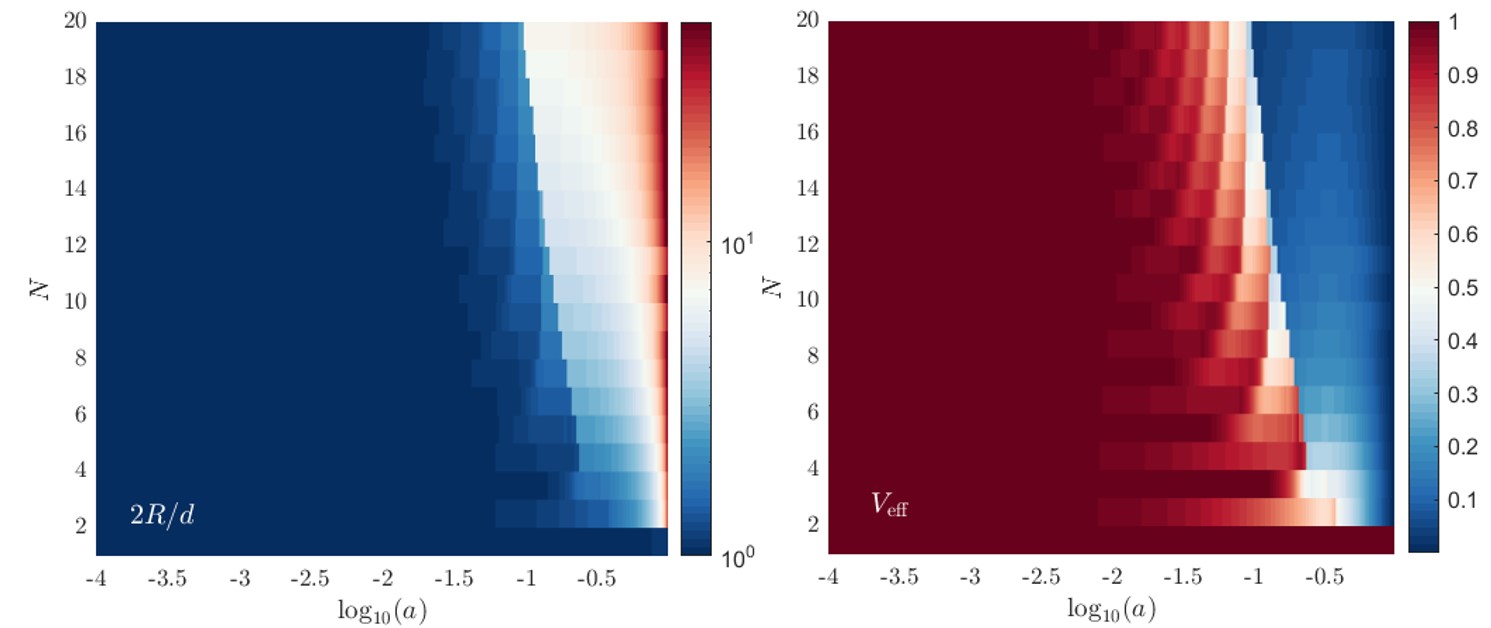}
    \caption{\textbf{Space of symmetric torus unknots for self-distance and effective volume.} The space of symmetric torus unknots spanned by $\log_{10}(a)$ and $N$ is presented. The left plot shows the effective aspect ratio due to the self-intersection limitation. The plot on the right considers the effective volume due to this limitation.}
    \label{fig:torusKnotMinD}
\end{figure*}

Coiling the axis greatly limits the effective aspect ratio of the resulting configuration (as well as the difficulty of the placement of coils to generate the field). This is because surfaces around the axis will tend to overlap. To measure this tendency more systematically we define a \textit{local self-distance}, $d$, at point $P$ along the curve $C$ as the smallest distance $d$ between $P$ and $\{Q\}$ on the plane $\pi$ normal to $C$ at $P$, where $\{Q\}$ is the set of intersections of $C$ with $\pi$ other than $P$. With this and using the expression for the symmetric torus unknot, we may then compute this parameter $d$ for all the configurations in $\{a, N\}$ space. To evaluate this distance we need first to select $\{Q\}$ by solving $(\mathbf{r}-\mathbf{r}_P)\cdot\mathbf{t}_P=0$ where $\mathbf{t}_P$ is the tangent to $C$ at $P$. The result is shown in Figure \ref{fig:torusKnotMinD}, were we present an effective aspect ratio as $2R/d$. \par
It is clear from Figure \ref{fig:torusKnotMinD} that the self-distance becomes severely limiting for largest $N$ and $a$. We also show what we call the \textit{effective volume}. The idea is to estimate the volume of the stellarator by considering it to have a uniform cross-section limited by $d$ and the length of the axis. So we define the effective volume as $V_\mathrm{eff}=Ld^2/8\pi$ (comparing the volume to that of a torus of minor radius 1). We observe that for small $a$, this limitation does not exist. The QA phase does not suffer from this problem and naturally allows for larger aspect ratios. Following $d$, the effective volume drops in the large $N$ and $a$ limit, a trend opposite to the writhe. Therefore, this will prevent saturating the contribution from the axis to the transform. The region where the effects are severe roughly corresponds to $a>1/N$, the asymptotic associated with the minimum curvature variation. Thus, from the QH phase, one would be interested in design candidates in the region $1/(a+N^2)<a<1/N$.

\section{Invariance of phase transitions} \label{sec:invarPhaseTrans}
In this appendix, we show the universality of the phase transitions that we found in Section IVC, Eqs.~(\ref{eqn:anharQAQHsurf}). The idea is to prove that these transitions prevail as one includes an arbitrary number of harmonics in the description of the magnetic axis, as well as allow for any choice of $Z_n$ with respect to $R_n$.\par
Let us start by constructing the expression for the curvature. The curvature can be written in the form $\kappa=\left|\mathrm{d}^2\mathbf{r}/\mathrm{d}\phi^2\times\mathrm{d}\mathbf{r}/\mathrm{d}\phi\right|/\left|\mathrm{d}\mathbf{r}/\mathrm{d}\phi\right|^{3/2}$, so we need to find, as a first step, an expression for the derivatives of $\mathbf{r}$, the position of the axis. Taking a general stellarator symmetric form for the axis, it is easy to show that, in cylindrical coordinates,
\begin{gather}
    \frac{\mathrm{d}\mathbf{r}}{\mathrm{d}\phi}=\sum_n\begin{pmatrix}
        - R_{n}n\sin n\phi \\
        R_n\cos n\phi \\
        Z_n n\cos n\phi
    \end{pmatrix},\\
    \frac{\mathrm{d}^2\mathbf{r}}{\mathrm{d}\phi^2}=-\sum_n\begin{pmatrix}
        R_{n}(1+n^2)\cos n\phi \\
        2R_n n\sin n\phi \\
        Z_n n^2 \sin n\phi
    \end{pmatrix}.
\end{gather} 
Here $R_0=1$, and the sum runs through all the harmonics describing the shape of the magnetic axis. \par
Now, the two transitions we discussed in Eq.~(\ref{eqn:anharQAQHsurf}) corresponded to a very particular set of curves. In fact, they corresponded to curves whose inflection points appear at $\phi=0$ and $\phi=\pi/N$, each describing one of the transitions. Thus, it is natural to evaluate these expressions we have obtained at $\phi=0,\pi/N$. Remarkably, the second derivative expression has a particularly simple form at these points: $\mathrm{d}^2\mathbf{r}/\mathrm{d}\phi^2(\phi=0)=-\sum_n s_n (1,0,0)$ and $\mathrm{d}^2\mathbf{r}/\mathrm{d}\phi^2(\phi=\pi/N)=-\sum_n (-1)^{n/N}s_n (1,0,0)$. Here we have used our universal measure $s_n=R_n(1+n^2)$. The vector has a single component, and as a result, the curvature at these points (from the expression for curvature) is proportional to these components. It then follows that the curvature at these points will vanish, respectively, if the conditions,
\begin{subequations}
\begin{gather}
    1+\sum_k s_{kN}=0, \\
    1+\sum_k (-1)^ks_{kN}=0,
\end{gather}\label{eqn:genAnharTrans}
\end{subequations}
are satisfied. These transitions define two hyperplanes in a space spanned by $\{s_n\}$. \par
These transitions are independent of the $Z_n$ shaping. In addition, they involve the universal measure $s_n$, which, as we have emphasized in the text, really appears to be the appropriate measure to use. Together with the transition points at $s_n=1$ (and all other $s_m=0$ for $m\neq n$), these two hyperplanes constitute invariant features of phase space in the sense that they do not change (in $s_n$ space) with $N$ nor $Z_n$ choices. A similar invariance does not appear to hold for other transitions, where additional geometric aspects are important.  

\bibliography{topSpaceQS}

\end{document}